# Sensitivity of a Point-Source-Interferometry-based Inertial Measurement Unit Employing Large Momentum Transfer and Launched Atoms

Jinyang Li[1], Timothy Kovachy[1], Jason Bonacum[2,3], and Selim M Shahriar[1,2]

[1]Department of Physics and Astronomy, Northwestern University, Evanston, IL 60208, USA
[2]Department of Electrical and Computer Engineering, Northwestern University, Evanston, IL 60208, USA
[3]Digital Optics Technologies, Rolling Meadows, IL 60008, USA

**Abstract**

We analyze theoretically the sensitivity of accelerometry and rotation sensing with a point source interferometer employing large momentum transfer (LMT) and present a design of an inertial measurement unit (IMU) that can measure rotation around and acceleration along each of the three axes. In this design, the launching technique is used to realize the LMT process without the need to physically change directions of the Raman pulses, thus significantly simplifying the apparatus. We also describe an explicit scheme for such an IMU.

## 1. Introduction

For inertial navigation, it is necessary to measure both acceleration and rotation. Multi-axis atom interferometers with such capabilities are being actively investigated [1, 2, 3, 4, 5, 6]. One challenge for a conventional atom interferometer is distinguishing between phase shifts produced by acceleration and rotation. The techniques used for this purpose include operating two or more interferometers where atoms move in opposite directions [1,2] and using a four-pulse interferometer protocol [1,3]. The counter-propagating technique [1,2] involves arguably an increased level of complexity, which may pose challenges for applications requiring portability. In the case of the four-pulse technique [1,3], the phase shift is proportional to the cross product of acceleration and angular velocity. Thus, in order to determine one of these measurands, the other must be accurately known. For measuring rotation on an Earth-bound platform, the acceleration due to gravity can be used as such a known parameter. However, for a space-borne platform, there does not exist such a known parameter. Although a bias acceleration or rotation could be applied to work as such a parameter, uncertainties in the applied biases would limit the accuracy of such a device significantly.

Arguably a more suitable method for distinguishing acceleration and rotation induced phase shifts is the point source interferometer (PSI) [7, 8, 9, 10, 11]. A one-axis PSI can be used for measuring acceleration along the primary axis, and two components of rotation along the axes orthogonal to the primary axis. In a conventional atomic interferometer, the signal is determined by measuring the total population of the atoms in a particular internal quantum state, and there exists a well-developed method for the sensitivity limit analysis for such a detection scheme [12, 13]. However, for a PSI, the information about the acceleration and the rate of rotation is distributed spatially over the whole atomic cloud. Specifically, the magnitude of rotation is inferred from the spacing of the spatial fringes formed by the atoms in one of two internal quantum states, and the direction of rotation can be gleaned from the orientations thereof. The acceleration, on the other hand, shifts the phase of the fringes. As such, an analysis of the sensitivity of rotation sensing and accelerometry for a PSI is very different from that of a conventional atomic interferometer. To the best of our knowledge, no systematic analysis of the theoretical sensitivity of a PSI signal exists in the literature, except for a heuristic model we presented in Ref. [14], but only for rotation sensing. In this paper, we present a systematic and rigorous analysis of both acceleration and rotation sensitivity of a PSI. We present a generalized expression for these sensitivities in terms of the signals measured by each pixel of a camera used for monitoring the distributed signal. We find that when the acceleration or rotation exceeds certain levels, it is possible to express the sensitivity in a closed form. Otherwise, the sensitivity can only be determined via a numerical evaluation of the general expression.

In order to determine the ultimate capability of a PSI, we also consider the case when Raman-transition-based large momentum transfer (LMT) [14, 15, 16, 17] is incorporated in order to enhance the sensitivity, for both rotation sensing and accelerometry. The way to implement LMT is to apply additional Raman $\pi$ pulses in opposite effective directions. The effective propagating direction of a pair of counterpropagating Raman beams is defined as the propagating direction of the higher-frequency Raman beam. Some of the additional $\pi$ pulses are effectively parallel to the original three Raman pulses and the rest are effectively antiparallel. One method for flipping the directions of the two Raman beams is to swap the polarizations of the two beams before a polarizing beam splitter (PBS) [16, 17]. To change the polarizations of the two beams rapidly without wasting optical power, a Pockels cell can be used [16]. A Pockels cell with a high extinction ratio needs to be driven by a voltage as high as several kilovolts. Furthermore, in this scheme, the two Raman beams must have orthogonal polarizations and thus cannot be two components in the output from an electro-optic modulator (EOM). This method for swapping the polarizations can increase the complexity of the system considerably. Another way to flip the directions of the Raman beams is to make use of the opposite Doppler shifts for Raman beam pairs with opposite effective directions [18]. The advantage of this method is that no external device is needed to swap the directions of the Raman beams. To induce a large Doppler shift for the Raman beam, a high bias velocity of the atoms is required. Usually, this bias velocity is induced by gravity. However, the free-fall approach has three drawbacks.

First, this approach can only impart a bias velocity in the vertical direction. Second, it takes a long time to accumulate a high bias velocity. Third, the atoms thermally expand during the free-fall process. In this paper, we propose to use moving-molasses-induced launching of atom to impart such a bias velocity for LMT.

The use of the launching technique for breaking the degeneracy between the transitions driven by the two pairs of Raman beams has been demonstrated previously [11]. What would be important is to show that this approach can be generalized in three dimensions with a single apparatus, in the context of implementing high-order LMT, since the ability to reverse the effective direction of the Raman beams via changing the modulation frequency would speed up the operation of the system. The advantage resulting from this speed-up of the LMT process would become more important for an inertial measurement unit (IMU), for measuring rotation around and acceleration along each of the three axes, since it would require the operation of three PSIs, either in parallel or serially. However, it is not a priori obvious how such a scheme can be generalized for three-axis operation in a manner that would not introduce other potential complications. As such, we present an explicit scheme, based on the use of $^{87}$Rb atoms, for implementing an IMU based on such a PSI, incorporating the LMT process as well as launching of atoms.

The rest of the paper is organized as follows. In Sec. 2, we present an analysis of the sensitivity of rotation sensing and accelerometry based on a signal image of a PSI. In Sec. 3, we analyze the effects of various experimental imperfections for the IMU. In Sec. 4, we present the general modality for an IMU employing point-source interferometry. In Sec. 5, we describe how the process of LMT is incorporated into the IMU. In Sec. 6, we present an explicit description for such an IMU, employing $^{87}$Rb atoms. Concluding discussions are presented in Sec. 7. The appendix shows the expression for phase shifts due to acceleration and rotation as functions of the order of LMT used.

## 2. Analysis of Acceleration and Rotation Sensitivity Based on a signal image of a PSI

To start with, we recall briefly how a conventional PSI operates on a terrestrial platform works. First, atoms are captured in a magneto-optic trap (MOT), and then undergo optical molasses cooling. Then, optical pumping and state selection is applied to place the atoms in the $m_F = 0$ Zeeman substate. Next, a bias magnetic field is turned on, and the atoms are allowed to fall freely for some time to gain a mean bias velocity so that different Raman resonant peaks resulting from a retroreflecting configuration can be separated. This step is followed by the application of Raman pulses in the vertical direction to induce atomic interference. Finally, an imaging beam is used to map the spatial distribution of atoms in one internal quantum state. This signal is then analyzed to infer the rates of rotation around the axes orthogonal to the vertical direction. It is also possible to infer the acceleration experienced along the vertical direction by analyzing the shift in the fringe pattern.

We next consider a PSI augmented with LMT. In **Figure 1**, we describe schematically the basic approach for LMT, for a PSI [14, 15, 16, 17]. Recalling briefly, the LMT scheme works as follows. The conventional sequence of three counter-propagating Raman pulses (denotes as $A_0$, $B_0$, and $C_0$) are augmented by additional counter-propagating $\pi$ pulses. For an LMT of order $n$, the process requires the application of $4n$ additional $\pi$ pulses. In **Figure 1**, we have illustrated the case for $n=2$. In this case, the first additional $\pi$ pulse ($A_1$) propagates in an effective direction opposite to that of the initial $\pi/2$ pulse ($A_0$). For the next $\pi$ pulse ($A_2$), the effective direction is opposite to the previous $\pi$ pulse ($A_1$). The effective directions of the rest of the additional $\pi$ pulses are also shown in **Figure 1**. It can also be seen that the pulse sequence is symmetric about pulse $B_0$. The pulse sequence for an arbitrary order of LMT can be obtained by generalizing this pattern. As shown in the **Appendix**, the acceleration-induced phase shift of an atom interferometer with LMT can be expressed as $\phi = \mathbf{k}_{\text{eff}} \cdot \mathbf{a} T \left( T + 2\sum_{j=1}^{n} T_j \right)$, and the rotation-induced phase shift as $\phi = (\mathbf{k}_{\text{eff}} \times \mathbf{\Omega}) \cdot \mathbf{r} \left( T + 2\sum_{j=1}^{n} T_j \right)$, where $\mathbf{k}_{\text{eff}}$ is the effective wavenumber of the counterpropagating Raman beams, whose absolute value is approximately $2k$, $\mathbf{a}$ is the linear acceleration, $\mathbf{\Omega}$ is the angular velocity, $\mathbf{r}$ is the displacement of the atoms from pulse $A_0$ to $C_0$. The values of the time intervals $T$ and $T_j$ are defined in **Figure 1**.

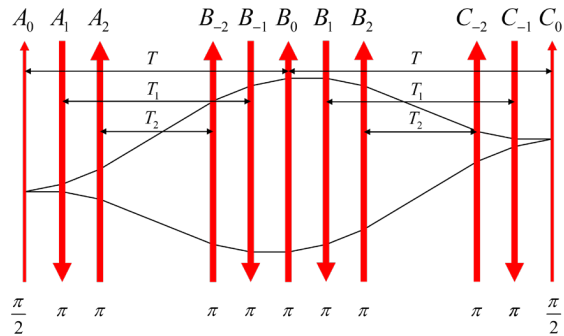

**Figure 1:** Illustration of the pulse sequence for the LMT protocol for $n=2$. For $n$th order LMT, $4n$ $\pi$ pulses are added. A pulse with an odd (even) subscript has an effective propagation direction opposite (identical) to that of the three original pulses.

Using the results shown above, we see that for a system employing $n$-

th order LMT, the ideal population distribution shown in the image should be

$$P(\mathbf{r}) = \frac{1}{2}\left[1 + \cos(\mathbf{k}_\Omega \cdot \mathbf{r} + \phi_a)\right] f(\mathbf{r}) \qquad (1)$$

where $\mathbf{k}_\Omega = \mathbf{k}_{eff} \times \mathbf{\Omega} T\left(T + \sum_{j=1}^{n} T_j\right)$, $\phi_a = \mathbf{k}_{eff} \cdot \mathbf{a} T\left(T + 2\sum_{j=1}^{n} T_j\right)$, and $f(\mathbf{r})$ is the profile of the final atomic cloud. It is easier to extract the angular velocity $\mathbf{\Omega}$ and the acceleration $\mathbf{a}$ in the Fourier domain. The Fourier transform of this image is

$$\tilde{P}(\mathbf{k}) = \frac{1}{2}\tilde{f}(\mathbf{k}) + \frac{1}{4}e^{i\phi_a}\tilde{f}(\mathbf{k} - \mathbf{k}_\Omega) + \frac{1}{4}e^{-i\phi_a}\tilde{f}(\mathbf{k} + \mathbf{k}_\Omega) \qquad (2)$$

where $\mathbf{k}$ is the conjugate variable of $\mathbf{r}$, and $\tilde{f}(\mathbf{k})$ is the two-dimensional Fourier transform (FT) of $f(\mathbf{r})$. There are three peaks in the Fourier domain. The two side peaks are located at $\pm\mathbf{k}_\Omega$, and the phases of the two side peaks are $\pm\phi_a$. Accordingly, the positions of the side peaks can be used to infer $\mathbf{\Omega}$, and the phases of the side peaks can be used to infer $\mathbf{a}$. From the expression of $\mathbf{k}_\Omega$ and $\phi_a$, it can be seen that only the component of $\mathbf{\Omega}$ perpendicular to $\mathbf{k}_{eff}$ and the component of $\mathbf{a}$ parallel to $\mathbf{k}_{eff}$ can be measured.

Physically, Eq. (2) can be understood as the convolution between the FT of $f(\mathbf{r})$, which is $\tilde{f}(\mathbf{k})$, and the FT of a uniformly spaced fringe pattern of infinite spatial extent. This pattern can be decomposed into a DC part, whose FT is a delta function centered at $\mathbf{k} = 0$ and an AC part, whose FT are two delta functions at $\mathbf{k} = \pm\mathbf{k}_\Omega$, multiplied by the phase factors. As such, the condition for the two peaks in Eqn. (2) to be clearly separable is that the extent of $\tilde{f}(\mathbf{k})$ must be smaller than $|\mathbf{k}_\Omega|$. This condition would be satisfied if the shape of the atomic cloud, $f(\mathbf{r})$, is smooth enough not to have variations on a scale smaller than the fringe spacings.

In a typical atomic sensor, such as an atomic clock, an atomic magnetometer, or a conventional atomic interferometer, the signal corresponds to the total number of atoms in a certain state (referred to as the detected state in what follows). In contrast, the signal of a PSI consists of the spatial distribution of the detected-state counts. Therefore, the well-established sensitivity analysis technique for the devices that only measure the total number of detected-state atoms cannot be applied to a PSI. As such, to the best of our knowledge, a systematic analysis of the rotation and acceleration sensitivity of a PSI, does not exist in the literature. The only exception is Ref. [14], in which we had developed only a heuristic expression for the rotation sensitivity of an

LMT augmented PSI; however, it did not include any estimation of the acceleration sensitivity.

Here we derive rigorously the expressions for the uncertainties in the measurement of acceleration and rotation using a PSI, for a generic situation that may or may not include augmentation via the LMT process. For specificity, we assume the situation where a camera with an array of pixels with identical sensitivities is used to capture a two-dimensional image of the atoms in the detected state. Without loss of generality, we consider the case where $\mathbf{k}_\Omega$ is in the $x$ direction, and the image is taken in the $x$-$y$ plane. In this case, it is not necessary to consider the 2D distribution of the signal. Instead, we first integrate the 2D image along the $y$ direction, thus producing an effective 1D signal along the $x$ direction. We note first that each atom can be either in or out of the detected state, so that the detection process is binary. As such, in a pixel labeled as $l$ in this 1D image, the number of detected-state atoms, $p_l$, follows a binomial distribution:

$$p_l \sim B\left[f(x_l), \frac{1}{2}(1+c\cos\phi)\right] \tag{3}$$

where $f(x_l)$ is the atom count in pixel $l$, $x_l$ is the $x$ coordinate of pixel $l$, $c$ is the contrast of the interference fringes, $\phi = k_\Omega x_l + \phi_a$, and $B(n,p)$ describes the binomial distribution $C_m^n p^m (1-p)^{n-m}$, where $m$ is the random variable. Therefore, it is easy to show that

$$\begin{aligned}\langle p_l \rangle &= \frac{1+c\cos\phi}{2}f(x_l) \\ \Delta p_l^2 &= \frac{1-c^2\cos^2\phi}{4}f(x_l)\end{aligned} \tag{4}$$

Although it is practically more convenient to analyze the signal in the Fourier domain, as illustrated in Sec. 6, a direct derivation of the measurement uncertainties in this domain involves a complicated propagation of uncertainty. Therefore, we consider the uncertainty derivation in the spatial domain directly. In this domain, one possible way to extract the values of $k_\Omega$ and $\phi_a$ is to fit the 1D image with the expected profile of $p_l$ via the process of least weighted squares regression. Explicitly, this corresponds to minimizing the value of the expression $\sum_{l=1}^{L}(p_l - \langle p_l \rangle)^2 / f(x_l)$, where $L$ is the number of pixels, by adjusting the parameters $k_\Omega$ and $\phi_a$. Therefore, the derivatives of this expression with respect to $k_\Omega$ and $\phi_a$ have null value:

$$\sum_{l=1}^{L} \frac{p_l - \langle p_l \rangle}{f(x_l)} \frac{\partial \langle p_l \rangle}{\partial k_\Omega} = 0$$

$$\sum_{l=1}^{L} \frac{p_l - \langle p_l \rangle}{f(x_l)} \frac{\partial \langle p_l \rangle}{\partial \phi_a} = 0 \tag{5}$$

Eq. (5) defines $k_\Omega$ and $\phi_a$ as implicit functions of $\{p_1, p_2, \cdots, p_L\}$. According to the rule for uncertainty propagation, the variances of $k_\Omega$ and $\phi_a$ can be expressed as

$$\Delta k_\Omega^2 = \left(\frac{\partial k_\Omega}{\partial p_1}\right)_A^2 \Delta p_1^2 + \left(\frac{\partial k_\Omega}{\partial p_2}\right)_A^2 \Delta p_2^2 + \cdots + \left(\frac{\partial k_\Omega}{\partial p_L}\right)_A^2 \Delta p_L^2$$

$$\Delta \phi_a^2 = \left(\frac{\partial \phi_a}{\partial p_1}\right)_A^2 \Delta p_1^2 + \left(\frac{\partial \phi_a}{\partial p_2}\right)_A^2 \Delta p_2^2 + \cdots + \left(\frac{\partial \phi_a}{\partial p_L}\right)_A^2 \Delta p_L^2 \tag{6}$$

where the subscript $A$ indicates that the derivative is evaluated at point $A$, and $A$ stands for the point $p_1 = \langle p_1 \rangle, p_2 = \langle p_2 \rangle, \cdots, p_L = \langle p_L \rangle$. Therefore, to calculate the variances of $k_\Omega$ and $\phi_a$, it is necessary to calculate their derivatives with respect to $p_1, p_2, \cdots, p_L$. Taking the derivative of Eq. (5) with respect to, for example, $p_1$, at point $A$, it is obtained that

$$\sum_{l=1}^{L} \left(\frac{\partial}{\partial p_1} \frac{p_l - \langle p_l \rangle}{f(x_l)}\right)_A \frac{\partial \langle p_l \rangle}{\partial k_\Omega} + \sum_{l=1}^{L} \left(\frac{p_l - \langle p_l \rangle}{f(x_l)}\right)_{p_l = \langle p_l \rangle} \left(\frac{\partial}{\partial p_1} \frac{\partial \langle p_l \rangle}{\partial k_\Omega}\right)_A$$

$$= \sum_{l=1}^{L} \left(\frac{\partial}{\partial p_1} \frac{p_l - \langle p_l \rangle}{f(x_l)}\right)_A \frac{\partial \langle p_l \rangle}{\partial k_\Omega} = 0$$

$$\sum_{l=1}^{L} \left(\frac{\partial}{\partial p_1} \frac{p_l - \langle p_l \rangle}{f(x_l)}\right)_A \frac{\partial \langle p_l \rangle}{\partial \phi_a} + \sum_{l=1}^{L} \left(\frac{p_l - \langle p_l \rangle}{f(x_l)}\right)_{p_l = \langle p_l \rangle} \left(\frac{\partial}{\partial p_1} \frac{\partial \langle p_l \rangle}{\partial \phi_a}\right)_A \tag{7}$$

$$= \sum_{l=1}^{L} \left(\frac{\partial}{\partial p_1} \frac{p_l - \langle p_l \rangle}{f(x_l)}\right)_A \frac{\partial \langle p_l \rangle}{\partial \phi_a} = 0$$

which follows that

$$\alpha \left(\frac{\partial k_\Omega}{\partial p_1}\right)_A + \beta \left(\frac{\partial \phi_a}{\partial p_1}\right)_A = \frac{1}{f(x_1)} \frac{\partial \langle p_1 \rangle}{\partial k_\Omega}$$

$$\beta \left(\frac{\partial k_\Omega}{\partial p_1}\right)_A + \gamma \left(\frac{\partial \phi_a}{\partial p_1}\right)_A = \frac{1}{f(x_1)} \frac{\partial \langle p_1 \rangle}{\partial \phi_a} \tag{8}$$

where

$$\alpha \equiv \sum_{l=1}^{L} \frac{1}{f(x_l)} \left( \frac{\partial \langle p_l \rangle}{\partial k_\Omega} \right)^2 = \frac{c^2}{4} \sum_{l=1}^{L} f(x_l) x_l^2 \sin^2 \phi$$

$$\beta \equiv \sum_{l=1}^{L} \frac{1}{f(x_l)} \frac{\partial \langle p_l \rangle}{\partial k_\Omega} \frac{\partial \langle p_l \rangle}{\partial \phi_a} = \frac{c^2}{4} \sum_{l=1}^{L} f(x_l) x_l \sin^2 \phi \quad (9)$$

$$\gamma \equiv \sum_{l=1}^{L} \frac{1}{f(x_l)} \left( \frac{\partial \langle p_l \rangle}{\partial \phi_a} \right)^2 = \frac{c^2}{4} \sum_{l=1}^{L} f(x_l) \sin^2 \phi$$

We now consider first the limit where the number of visible fringes is large enough so that the peaks shown in Eq. (2) are separated to the extent that the overlaps between the spectral (spatial-frequency) distributions of the neighboring peaks are negligible. In this case, the coefficients in Eq. (9) can be approximated as

$$\alpha \approx \frac{c^2}{4} \langle \sin^2 \phi \rangle \sum_{l=1}^{L} f(x_l) x_l^2 = \frac{c^2 N \sigma_f^2}{8}$$

$$\beta \approx \frac{c^2}{4} \langle \sin^2 \phi \rangle \sum_{l=1}^{L} f(x_l) x_l = 0 \quad (10)$$

$$\gamma \approx \frac{c^2}{4} \langle \sin^2 \phi \rangle \sum_{l=1}^{L} f(x_l) = \frac{c^2 N}{8}$$

Then Eq. (8) can be simplified as

$$\left( \frac{\partial k_\Omega}{\partial p_1} \right)_A = \frac{8 \partial \langle p_1 \rangle / \partial k_\Omega}{N c \sigma_f^2 f(x_1)} = -\frac{4 x_1 \sin \phi}{N c \sigma_f^2}$$

$$\left( \frac{\partial \phi_a}{\partial p_1} \right)_A = \frac{8 \partial \langle p_1 \rangle / \partial \phi_a}{N c f(x_1)} = -\frac{4 \sin \phi}{N c} \quad (11)$$

Substituting Eq. (11) into Eq. (6), it can be obtained that

$$\Delta k_\Omega^2 \approx \frac{4 \sum_{l=1}^{L} f(x_l) x_l^2}{N^2 c^2 \sigma_f^4} \langle \sin^2 \phi (1 - c^2 \cos^2 \phi) \rangle = \frac{4 - c^2}{2 N c^2 \sigma_f^2}$$

$$\Delta \phi_a^2 \approx \frac{4 \sum_{l=1}^{L} f(x_l)}{N^2 c^2} \langle \sin^2 \phi (1 - c^2 \cos \phi) \rangle = \frac{4 - c^2}{2 N c^2} \quad (12)$$

Therefore, the sensitivity limits can be expressed as:

$$\Delta\phi_a^{-1} = c\sqrt{\frac{2N}{4-c^2}} \approx c\sqrt{\frac{N}{2}}$$
$$\Delta k_\Omega^{-1} = c\sigma_f\sqrt{\frac{2N}{4-c^2}} \approx c\sigma_f\sqrt{\frac{N}{2}} \quad (13)$$

Substituting the expressions for $k_\Omega$ and $\phi_a$, the precision of the acceleration and the angular velocity can be expressed as:

$$\Delta a^{-1} = k_t T^2 c\sqrt{\frac{N}{2}}; \quad \Delta\Omega^{-1} = k_t T c\sigma_f\sqrt{\frac{N}{2}} \quad (14)$$

where $k_t$ is the momentum transfer divided by $\hbar$.

When we take into account the number of measurement cycles that are repeated within a certain period of time $\tau$, the measurement-time-dependent sensitivities can be expressed as:

$$\Delta a^{-1} = k_t T^2 c\sqrt{\frac{N\tau}{2\tau_0}}; \quad \Delta\Omega^{-1} = k_t T c\sigma_f\sqrt{\frac{N\tau}{2\tau_0}} \quad (15)$$

where $\tau_0$ is the single-shot duration, including any dead-time between each cycle. For a specific set of parameters, namely $k_t = 10 k_{eff}$, $T = 20$ ms, $c = 0.5$, $N = 10^6$, $\tau_0 = 1$ s, and $\sigma_f = 1$ mm, measurement uncertainties corresponding to these sensitivities are calculated to be $\Delta\Omega\sqrt{\tau} = 0.88$ µrad·s$^{-1}$·Hz$^{-1/2}$ and $\Delta a\sqrt{\tau} = 4.5$ nano-$g$·Hz$^{-1/2}$.

In cases where the spectral (spatial-frequency) distributions of the neighboring peaks shown in Eq. (2) still overlap significantly, numerical simulations based on Eqs. (6), (8), and (9) are needed to find the exact sensitivities because there are no simple analytical expressions for them, keeping in mind that the sensitivities are also proportional to the square root of the repetition number of the measurement cycles. Nevertheless, the analytical results shown here serve as important indicators for estimating the achievable sensitivities. The sensitivity analysis presented above applies to each axis of the 3-axis IMU, since it operates in a sequential mode. It should be noted that the expression of rotation sensitivity shown in the right part of Eq. (14) agrees with the heuristic expression for the same given in Ref. [14]. However, the analysis presented above applies more broadly to situations where the constraint used in deriving Eq. (14) is not valid. Furthermore, it should be noted again that no analysis of acceleration sensitivity, even a heuristic one, was given in Ref. [14].

To show how the sensitivity converges to the expressions shown in Eq.

(13) as $k_\Omega$ increases, the accelerometry sensitivity versus $k_\Omega \sigma_f$ for Gaussian $f(x)$ is plotted in **Figure 2** for $\phi_a = 0$ and $\phi_a = \pi/2$, using numerical simulations based on Eqs. (6), (8), and (9). It can be seen that when $k_\Omega \sigma_f$ is close to zero, the sensitivity depends on $\phi_a$, and that as $k_\Omega \sigma_f$ increases to $\pi$, the sensitivity converges to the expressions shown in Eq. (13) regardless of the value of $\phi_a$.

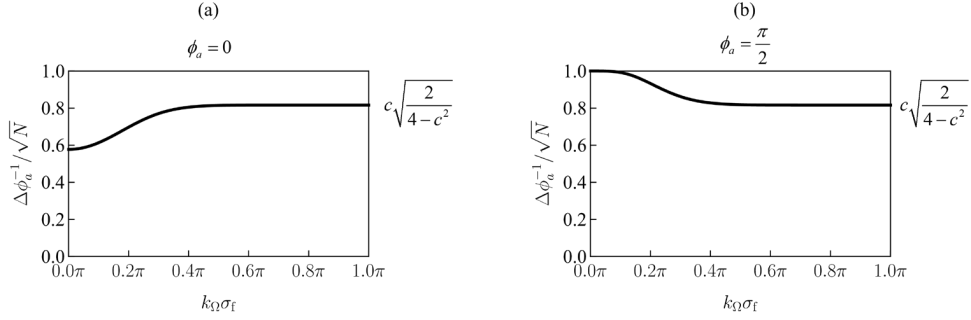

**Figure 2:** Accelerometry sensitivity of a PSI as a function of $k_\Omega \sigma_f$ for Gaussian $f(x)$ when (a) $\phi_a = 0$ and $\phi_a = \pi/2$. When $k_\Omega \sigma_f$ is close to zero, the sensitivity depends on $\phi_a$. As $k_\Omega \sigma_f$ increases to about $\pi$, the sensitivity converges to the analytical expressions.

It should also be noted that measurement of acceleration with any atomic interferometer, including the PSI, becomes ambiguous when the phase shift due to acceleration exceeds the value of $2\pi$. This is not a problem for a platform for which the acceleration is expected to be small enough so that the phase shift would be much less than $2\pi$. For platforms that may undergo much larger acceleration, it would be necessary to make use of a coarse-grained auxiliary accelerometer for disambiguation of the signal [19]. In the case of the IMU, three such auxiliary accelerometers would be required.

## 3. Experimental imperfections affecting the PSI sensitivity

From the analytical results presented above, we see that the parameters that affect the sensitivities are as follows: the fringe contrast ($c$), the order of LMT ($k_t$), the duration of the Raman pulse sequence ($2T$), the number of atoms interrogated in one shot ($N$), the single-shot duration ($\tau_0$), and the final size of the atomic cloud ($\sigma_f$), which only affects the rotation measurement sensitivity. In practice, the fringe contrast $c$ depends on some other parameters due to various imperfections, which can be divided into two categories. The first category pertains to effects that prevent the transition efficiency of the Raman pulses from reaching unity. These include Doppler-shift-induced detuning, spatial inhomogeneity and temporal fluctuation of the Raman beam intensity, and spontaneous emission. The second category pertains to other imperfections that are unrelated to the Raman pulse transition efficiency. These include a finite initial size of the atomic cloud and the imaging noise. We next consider these two categories in sequence.

An important property of the imperfections in the first category is that they result in a reduction of $c$ as the order of LMT increases. Therefore, generally, there exists an optimal order of LMT that maximizes the sensitivity for a given set of operating parameters. In Ref. [14], we studied optimization of the LMT order when Doppler-shift-induced detuning and spontaneous emission are taken into account. In Figure 12 of Ref. [14], the sensitivity enhancement as a function of the order of LMT is plotted for a set of discrete Raman beam intensities. However, it is also instructive to study how the maximum sensitivity enhancement varies as a function of the Raman beam intensity. We have carried out such a study, based on the analytical result presented in Ref. [14]. Some results in Ref. [14] are summarized as follows. The two-photon Rabi frequency can be expressed as $\Omega^2/2\Delta$, where $\Omega$ is the single-photon Rabi frequency and $\Delta$ is the single-photon detuning, and the effective spontaneous decay rate can be expressed $\Gamma\Omega^2/4\Delta$. Therefore, for a fixed $\Omega$, a larger $\Delta$ decreases both the two-photon Rabi frequency and the spontaneous decay rate. As such, there exists an optimal $\Delta$ that maximizes the signal contrast, and this optimal $\Delta$ depends on $\Omega$. If the optimal detuning is adopted, the optimal order of LMT and the corresponding maximum sensitivity enhancement as a function of the Raman beam intensity are shown in **Figure 3**. It should be kept in mind that for each Raman beam intensity, the maximum enhancement factor requires a different number of LMT pulses. These considerations have to be taken into account in the operation of the IMU.

There are some methods to suppress the effect of the large relative Doppler shift between the two interferometer arms when the order of LMT is high. These methods include addressing each arm individually [20], which has been applied to Bragg transitions, and using Floquet pulses [21], which has been applied to single-photon transitions. In principle, these methods can also be applied to Raman transitions. However, detailed theoretical modelling is needed to see how well these techniques would work in this case of Raman transitions, and will be carried out by us in the near future.

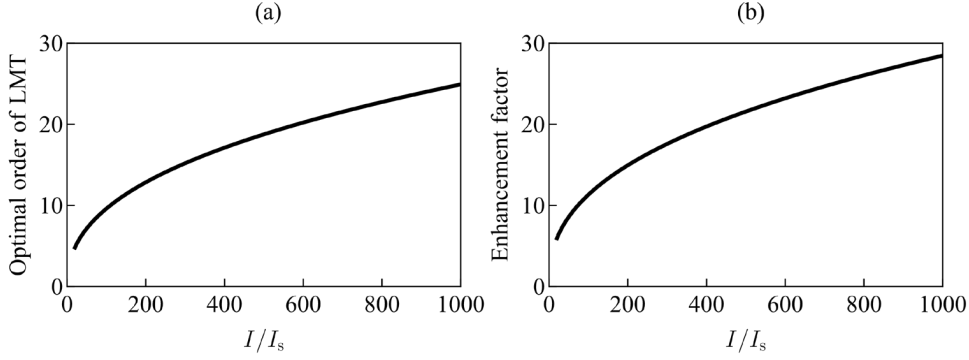

**Figure 3:** (a) Optimal order of LMT as a function the Raman beam intensity. For $n$th order of LMT, $k_t = (2n+1)k_{\text{eff}}$. For the Raman transition $F=1, m_F = 0 \rightarrow F = 2, m_F = 0$, $I_s = 5.01 \text{ mW} \cdot \text{cm}^{-2}$. (b) LMT-induced sensitivity enhancement factor as a function of the light intensity.

For the second category, we note first that the effect of the imperfections related to imaging would depend on the experimental details of any particular implementation of the PSI. In contrast, the effect of the finite initial size of the atomic cloud is more fundamental. As such, we consider here the effect of this imperfection in some details. For a PSI where the initial radius of the atomic cloud $\sigma_0 \neq 0$, a larger $T$ does not only increases the phase shift, but also makes the actual sensitivity closer to the ideal one by increasing the ratio $\sigma_f / \sigma_0$, where $\sigma_f$ is the final radius of the atomic cloud. However, a larger $T$ increases the drifting and falling distances of the atoms. There is a trade-off between the effect of $\sigma_f / \sigma_0$ and the apparatus size. For convenience, we define the parameter $b \equiv 1 - \sigma_0^2 / \sigma_f^2$, which is called the fringe broadening factor since the actual wavenumber of the fringes when $\sigma_0 \neq 0$ is given by $k'_\Omega = b k_\Omega$. This factor also reduces the contrast according to the following relation: $c' = c \exp\left[-k_\Omega^2 b(1-b)\right]$ [7,14]. Therefore, the ratio of the actual sensitivity to the ideal one can be expressed as $b \exp\left[-k_\Omega^2 b(1-b)\right]$. This ratio is plotted in **Figure 4** for $\sigma_0 = 0.2$ mm and $k_\Omega = \pi/2 \text{ cm}^{-1}$, when the temperature of the atoms is 6 μK. It can be seen that this ratio can exceed 80% when $T = 20$ ms. With $T = 20$ ms (corresponding to an expansion time of 40 ms), and assuming a bias velocity of, for example, $1 \text{ m} \cdot \text{s}^{-1}$, the distance traveled by the atoms during this process would be about 4 cm. The bias velocity must induce a relative Doppler shift much higher than the width of the Doppler broadening, as well as the Doppler shift corresponding to the momentum transfer. For $n$th order of LMT, the Doppler shift resulting from the momentum transfer can be estimated as $n \times 30$ kHz.

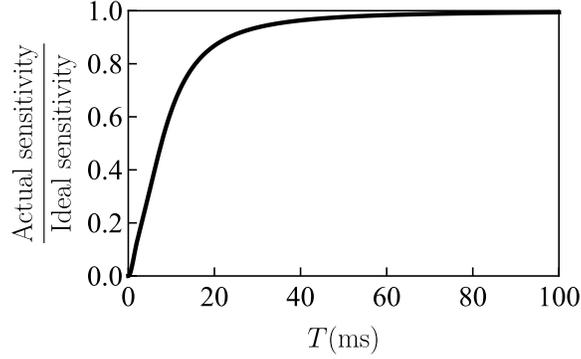

**Figure 4:** Ratio of the actual sensitivity to the ideal sensitivity for $\sigma_0 = 0.2$ mm and $k_\Omega = \pi/2$ cm$^{-1}$, when the finite initial size of the atomic cloud is considered. The temperature of the atoms is set at 6 µK.

We finally consider systematic errors, which are characterized by the property that they cannot be corrected by more repetitions of the measurement. Therefore, careful analysis and calibration are needed to alleviate the effect of such errors. Sources of systematic errors encompass a wide range and strongly depend on the specific experimental system. Here, as an example, we consider the effect of magnetic field gradient. In most atomic sensors, the effect of magnetic fields and gradients thereof can be suppressed to an arbitrary degree by employing magnetic shielding, and avoiding the presence of any strong magnetic field nearby. However, for the particular design of the PSI we have proposed here, the unavoidable use of a Faraday rotator not too far from the atoms implies that the effect of magnetic field gradient may be more substantial than it is for other atomic sensors. This would be true even if a magnetic shield is employed around the atoms, since one must allow openings in the shield for the optical beams to pass through.

Although only the $m_F = 0$ Zeeman substates are involved in the interference process, they are not completely resistant to external magnetic fields, due to higher order Zeeman shifts. If the atoms move through a spatially varying magnetic field, the energy gap between the two $m_F = 0$ Zeeman substates also vary, which can potentially induce a phase shift. According to the Breit–Rabi formula [22,23], the $F = 1(2), m_F = 0$ Zeeman substate is subject to a second order Zeeman shift of $\pm\Delta_Z = \pm(g_S \mu_B B)^2/4\Delta_{HFS}$. Therefore, the relative second order Zeeman shift between the two $m_F = 0$ Zeeman substates is shown to be $2\Delta_Z$, where $g_S \approx 2$ is the g-factor of the electron spin, $\mu_B$ is the Bohr magneton, $B$ is the magnetic field magnitude, and $\Delta_{HFS}$ is the hyperfine splitting of the ground state.

To estimate the phase shift caused by this second order Zeeman effect, we consider an example case where an atom is exposed to a magnetic field of 3 G in the first half of the Raman pulse sequence, but moves to a region with a magnetic field of 3.01 G in the second half. These values of the magnetic field and the gradient thereof are merely examples, based on what might reasonably be expected in the presence of a Faraday rotator in the path of the Raman beams. For a physical implementation of the PSI employing the Faraday rotator, one has to measure the actual magnetic field and the gradient thereof, and scale the results of the analysis presented below accordingly.

For these values of the magnetic field and its gradient, the energy gap between the two $m_F = 0$ Zeeman substates differs by 35 Hz between the first half and the second half of the Raman pulse sequence. For $T = 20$ ms, the extra phase shift would be about $1.4\pi$. Another effect of the magnetic field gradient is a force acting on the atoms resulting from the gradient of the second order Zeeman shift. The $F = 1(2), m_F = 0$ Zeeman substate is a high (low) field seeking state. The force can be calculated to be $\hbar d\Delta_Z/dz = (dB/dz)(\hbar g_S^2 \mu_B^2 B/2\Delta_{HFS})$, which is about $1.1 \times 10^{-23}$ dyn when $B = 3$ G and $dB/dz = 1$ G·cm$^{-1}$. This force causes an acceleration of about 81 micro-$g$ for $^{87}$Rb atoms. Therefore, it would be very difficult to measure the absolute acceleration directly from the phase shift. Even for measuring a relative acceleration, strong magnetic shielding may be needed to decrease the uncertainty due to the magnetic field distribution.

4. **Conceptual modality of an inertial measurement unit employing point source interferometry**

To realize a PSI-based IMU suitable for an arbitrary platform that may or may not be terrestrial, we present a concrete scheme, as illustrated in **Figure 5**. To simplify the discussion, we defer to the next section the technical details necessary for augmenting this system with LMT for enhancing sensitivity. A single MOT inside a spherical vacuum chamber is to be used for such an IMU. One pair of beams for the MOT, denoted as MOT 1 and MOT 1′, is applied along the *x* axis, as shown in the top panel of **Figure 5**. The other two pairs are applied in the *y-z* plane (MOT 2, MOT 2′, MOT 3, MOT 3′), as shown in the bottom panel of **Figure 5**. Additional beams are used to realize three different PSIs, orthogonal to one another. For each PSI, a pair of Raman beams and an imaging beam are first combined and sent through an optical fiber.

Consider first the case of one of these PSIs, denoted as PSI-1. The propagation axis for PSI-1 is along the *z* axis (PSI 1), as shown in the top panel of **Figure 5**. We assume that absorptive instead of fluorescence imaging is to be used, so that the cameras do not need to be placed very close to the atomic cloud. The imaging system for PSI-1 is also shown in the top panel of **Figure**

5. The Raman beams have the polarization of $(|x\rangle+|y\rangle)/\sqrt{2}$, where $|x(y)\rangle$ denotes the linear polarization along the $x(y)$ axis. The imaging beam has the polarization of $(|x\rangle-|y\rangle)/\sqrt{2}$. The Raman beams and imaging beams are aligned with the fast and the slow axis of the optical fiber. After they pass through the upper quarter waveplate, the Raman beams become $\sigma^+$-polarized and the imaging beam $\sigma^-$-polarized. The fast axis of the lower quarter waveplate is aligned with the slow axis of the upper one, so that the beams are all restored to their original linear polarizations. Then a Faraday rotator turns the polarization of the Raman beams to $|y\rangle$ and the imaging beam to $|x\rangle$. Given the state-of-the-art technology, the aperture of the Faraday rotator will not be sufficiently large for the Raman beams. To circumvent this problem, a pair of lenses can be used (not shown) to reduce the size of the optical beams before entering the Faraday rotator. Of course, this would introduce a scale factor between the actual angle of rotation experienced by the reflected beams at the atomic cloud and the angle of rotation of the retro-reflecting mirror, and this factor has to be taken into account when using such a rotation to calibrate the rotation sensitivity of the PSI. [24].

The Raman beams are transmitted through a PBS and the imaging beam is reflected by the PBS to the camera. The retroreflected Raman beams have the polarization of $(|x\rangle-|y\rangle)/\sqrt{2}$ after they pass through the Faraday rotator for the second time. In this way, after they pass through the lower quarter waveplate, they become $\sigma^+$-polarized again. It should be noted that the Faraday rotator is necessary for restoring the polarization after the Raman beam is retroreflected. We assume that the reflections from the windows can be ignored due to either antireflection coating or a slight angle between the Raman beams and the normal direction of the windows. In what follows, we will use the following notations for different components of the Raman beams. Specifically, the Raman beams propagating in the $+\hat{z}(-\hat{z})$ direction will be denoted as $R_1^{+z}\left(R_1^{-z}\right)$ and $R_2^{-z}\left(R_2^{+z}\right)$. The other two PSI-s (PSI-2 and PSI-3) have propagation axes in the $x$-$y$ plane, as shown in the bottom panel of **Figure 5**. These two PSIs are configured in a manner identical to that of PSI-1.

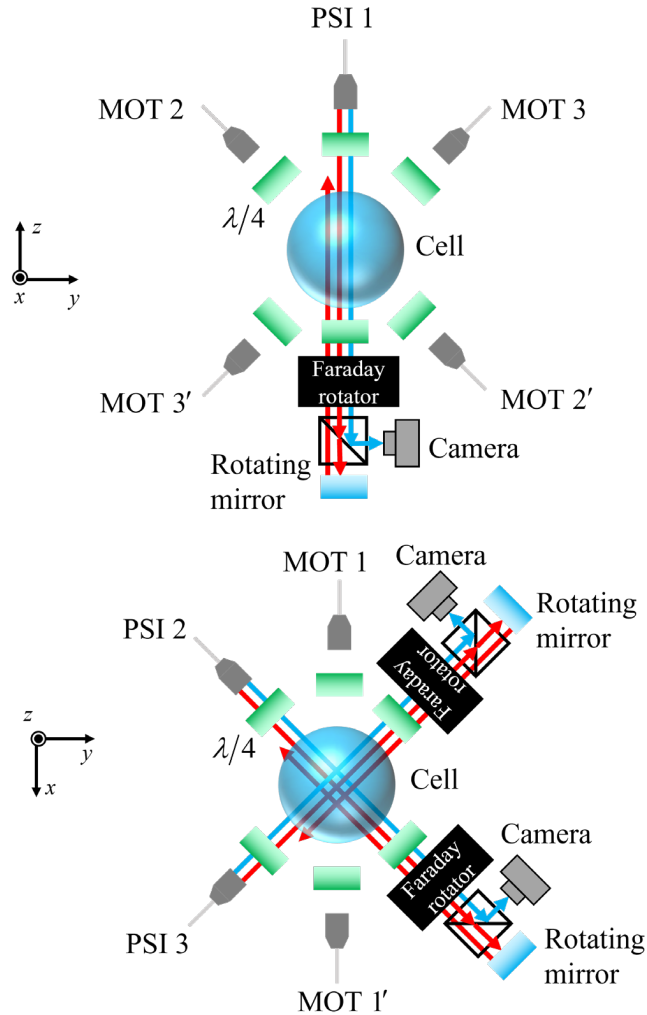

**Figure 5**: Schematic illustration of a three-dimensional PSI apparatus. The diagram on the top shows a cross-sectional view in the *y-z* plane. Here, MOT 2 and MOT 2′ are a pair of counterpropagating beams for the magneto-optic trap (MOT). So are MOT 3 and MOT 3′. The remaining pair needed for the MOT is not shown in this view but rather in the diagram at the bottom, since they propagate along the *x* direction. PSI 1 represents all the beams used for the PSI process along the *z* axis. Specifically, the red arrows represent the Raman beams, the blue ones represent the imaging beams. The green rectangles represent quarter waveplates and the black boxes represent Faraday rotators. The quarter waveplates and the Faraday rotator are configured in such a way that the Raman beams pass through the PBS and get retroreflected while the imaging beam is reflected by the PBS to the camera. The Faraday rotator is necessary since the incident and the retroreflected Raman beams are both $\sigma^+$-polarized in the vacuum chamber. The diagram on the bottom shows a cross-sectional view in the *x-y* plane, without showing the beams that are already illustrated in the top diagram. Here PSI 2 and PSI 3 represent the other two sets of beams (Raman beams and imaging beams) for the third PSI process in the *x-y* plane. The waveplates for these PSIs are configured in a manner analogous to that described for PSI 1.

To generate a bias velocity along each of the three PSI axes, we propose the use of moving optical molasses [2,28]. In what follows, this operation is referred to as launching. For each PSI, the moving optical molasses is created by reducing the frequencies of some MOT beams. Consider first the case of PSI-1. For this PSI, we have to reduce the frequencies of MOT 2′ and MOT 3′. This will produce a velocity along the MOT2-MOT 2′ axis and the MOT 3-MOT 3′ axis. Two of the velocity components will cancel each other, while the other two will add to each other, thus generating a net velocity along the z-axis. Using similar arguments, it is easy to determine the MOT beams whose frequencies have to be reduced for the other two PSIs. The complete list for the MOT beams whose frequencies need to be reduced for each of the three PSIs is shown in **Table 1**. It should be noted that under this process the atoms are adiabatically dragged along by the moving dipole force potentials produced by the molasses beam. As such, the velocity of the atoms at the start of the interferometry process would be essentially independent of any ambient acceleration, such as the one due to gravity in a terrestrial apparatus, or those due to vibrations.

**Table 1**: Relation between the moving direction of the molasses and the frequency configuration of the beams.

| Moving direction of the molasses | Beams whose frequencies are reduced |
|---|---|
| Propagation axis for PSI-1 | MOT 2′, MOT 3′ |
| Propagation axis for PSI-2 | MOT 1′, MOT 2′, MOT 3 |
| Propagation axis for PSI-3 | MOT 1, MOT 2′, MOT 3 |

We note that both co-propagating and counter-propagating Raman transitions can be excited by the Raman beams, as illustrated schematically in **Figure 6**, using the case of the D2 manifold in $^{87}$Rb for concreteness. In **Figure 6**(a), we show the process that would generate co-propagating Raman excitations, in both $+\hat{z}$ and $-\hat{z}$ directions. It should be noted that for a co-propagating Raman excitation, the difference in the Doppler shifts is very small, and can be ignored. In this diagram, we have considered only the Raman transitions that couple the $F=1, m_F=0$ and $F=2, m_F=0$ Zeeman sublevels, under the assumption that the atoms in the other Zeeman sublevels would be blown away prior to the PSI operation, as discussed later. In **Figure 6**(b) we show one possible counter-propagating Raman transition, and in **Figure 6**(c) we show the process that would generate the other counter-propagating Raman excitation. As indicated in these figures, these two types of counter-propagating Raman transitions would have opposite Doppler shifts. For a given spread of velocities in the atoms, the velocity imparted by the launching process has to be large enough to distinguish between these three types of Raman transitions [25].

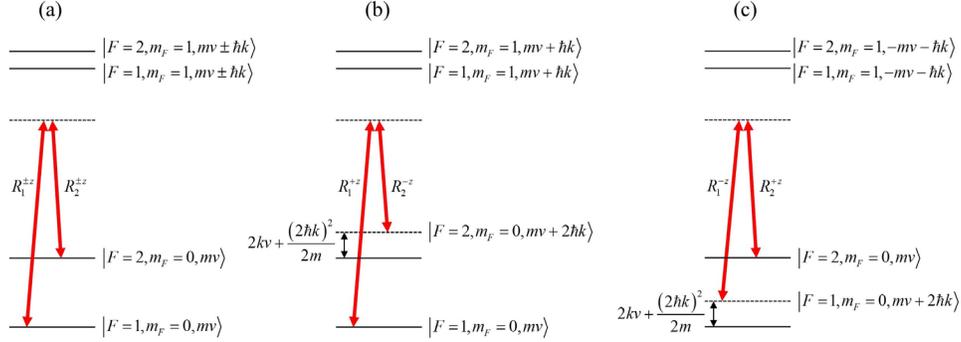

**Figure 6**: Coupling between $F=1, m_F=0$ and $F=2, m_F=0$ Zeeman ground states of $^{87}$Rb induced by copropagating and counterpropagating Raman beams along the $z$ axis. Here, $v$ is the amplitude of the atomic bias velocity in the $z$ direction. The bias velocity is assumed to be in the $z$ direction. $R_{1(2)}^{\pm z}$ represents the higher (lower) frequency Raman beam in the $\pm z$ direction. The wavenumbers of both Raman beams are denoted as $k$, and the small wavenumber difference between them is neglected. In the designations of the quantum states, the last element indicates the momentum in the z-direction. The transitions driven by three pairs of Raman beams have three different resonant frequencies in the presence of a bias velocity. (a) Coupling induced by copropagating Raman beams. The transition frequency is simply the hyperfine splitting. (b) Coupling induced by $R_1^{+z}$ and $R_2^{-z}$. The transition frequency is shifted up by $2kv + (2\hbar k)^2/2m$, where $2kv$ is Doppler shift and $(2\hbar k)^2/2m$ is the recoil frequency. (c) Coupling induced by $R_1^{-z}$ and $R_2^{+z}$. The transition frequency is shifted down by $2kv + (2\hbar k)^2/2m$.

The launching technique has two advantages compared to the case of free fall employed in a conventional, one-dimensional PSI. The first advantage is that this technique makes it possible to make use of Doppler shifts to discriminate between the three types of Raman transitions discussed earlier for any orientation of the PSI, including the case when it is horizontal. Of course, for a horizontal PSI, the effect of falling of the atoms transverse to the Raman beams needs to be taken into account, if operating the IMU terrestrially or in a near-Earth platform. Specifically, the value of Raman pulse sequence time, $2T$, needs be chosen in a manner that balances the trade-off between the effect of $\sigma_f/\sigma_0$ shown in **Figure 4** and the falling distance. As described earlier, a reasonable value of the sequence time is $2T = 40$ ms, which yields a value of ~80% for the ratio of the actual sensitivity to the ideal one. For this choice of the sequence time, the falling distance would be about 8 mm. In order to ensure that such a drop does not produce a significant change in the Raman Rabi frequency, it would be necessary to ensure that the radius of each of the Raman beams be at least ~4 cm. Alternatively, one can choose to use a smaller value of the sequence time, at the expense of reducing the ratio of the actual sensitivity to the ideal one.

The second advantage is that the atomic cloud will expand less when gaining the bias velocity, for two reasons. First, the acceleration induced by a moving optical molasses can be much higher than gravitational acceleration, and shorten the time of expansion. For example, in Ref. [26], atoms are accelerated to 2.6 m·s$^{-1}$ within 0.5 ms. To impart such a bias velocity with gravity, 270 ms is needed. Second, the atomic cloud expands much more slowly in an optical molasses field. For example, in Ref. [27], the $e^{-2}$ radius of the atomic cloud increases by 0.2 mm within 100 ms in the molasses field. The atoms released from this optical molasses are at a temperature of 6 μK. If the free-fall approach were to be used for such an ensemble of atoms, the radius would increase by 4.8 mm within 100 ms. If the atomic cloud remains relatively small at the application of the first Raman pulse, with a fixed constraint on the final size of the atomic cloud, the dark period duration $T$ can be longer, which increases the phase shift.

The state selection can be implemented with the counterpropagating Raman pulse [26,28,29] and a blow-away pulse, realized using the imaging beam for each PSI. In this way, the atoms are not only concentrated in the $F = 1, m_F = 0$ Zeeman substates, but also at a lower temperature longitudinally, due to the high Doppler sensitivity of the counterpropagating Raman transition. The transverse temperature of the atoms would remain unaffected by this process.

## 5. Incorporation of large momentum transfer in the PSI-based IMU

We next discuss the method to implement LMT with such a PSI-based IMU. It is evident that the LMT process described in **Figure 1** can be applied to each of the three PSIs. In Ref. [14], we presented a detailed analysis of how the enhancement in sensitivity varies with increasing order of LMT. The phase shift increases with the order of LMT according to the expression derived above, causing narrowing of the interference fringes. However, as the order of LMT increases, the fringe amplitude decreases, due to increasing Doppler shift and spontaneous emission. Therefore, there exists an optimal order of LMT that maximizes the sensitivity. The optimal order and the maximum sensitivity increase monotonically with the intensity of the Raman beams. The results of this analysis remain valid for each of the three PSIs in the IMU.

It should be noted that, for the modality described in **Figure 5**, no additional optical beams or hardware is needed for implementing the LMT process, for any order. To see why, note that for the LMT protocol illustrated in **Figure 7**(a), the effective propagation directions of the (counter-propagating)

Raman beams need to be swapped for some $\pi$ pulses. As illustrated earlier in **Figure 5**, there already exists two pairs of counterpropagating Raman beams in opposite effective directions. For concreteness, we assume that the two frequencies for the Raman excitation are generated with an electro-optical modulator (EOM) driven by a voltage-controlled oscillator (VCO), as illustrated in more details in Sec. 6. When the atoms move with a bias velocity parallel to the Raman beams, the resonant frequencies of these two pairs of Raman beams are different due to the Doppler shift. Therefore, by tuning the frequency of the VCO, it is possible to make the atoms experience Raman transitions due to only one of these counter-propagating Raman beams, and not the other. Thus, the propagation directions of the Raman beams can be effectively swapped by simply tuning the frequency of the VCO. The timing of the VCO frequency necessary for the $n=2$ LMT protocol is illustrated in **Figure 7**(b).

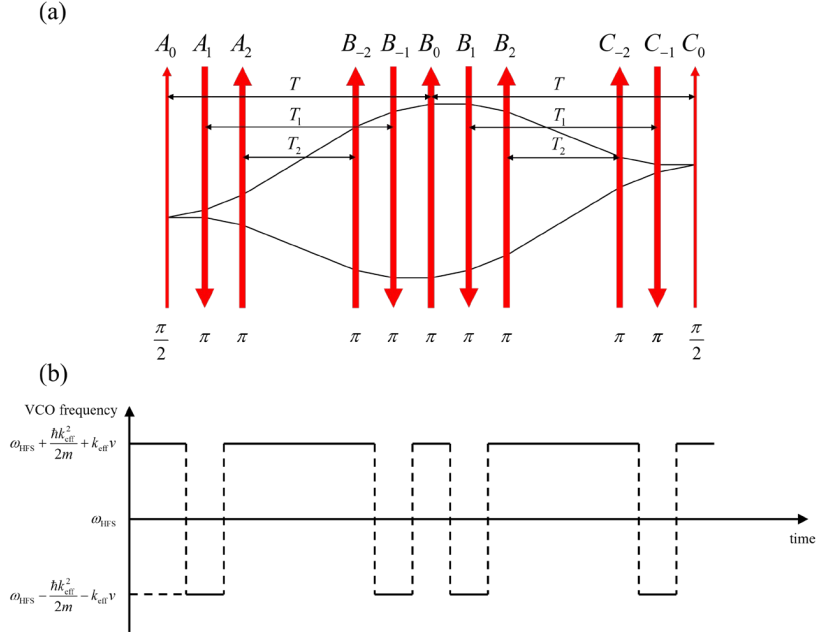

**Figure 7:** (a) Illustration of the pulse sequence for the LMT protocol for $n=2$. (b) Timing of the VCO frequency for this protocol. The VCO determines the relative frequency between the two Raman beams. When the VCO is tuned to $\omega_{\text{HFS}} + (\hbar k_{\text{eff}})^2/2m + kv$, $R_1^{+z}$ and $R_2^{-z}$ take effect. When the VCO is tuned to $\omega_{\text{HFS}} - (\hbar k_{\text{eff}})^2/2m - kv$, $R_1^{-z}$ and $R_2^{+z}$ take effect. See text for details.

## 6. Specific Layout of a PSI-based IMU using $^{87}$Rb atoms

For concreteness, a possible implementation of a PSI-based IMU employing $^{87}$Rb is shown in **Figure 8**. **Figure 8**(a) shows the layout of the optical beams. Only one laser is needed in this implementation, assuming it has high enough output power. For example, Toptica TAPro with an output of 4 W should work for such a design. The laser frequency is locked to the cross-over saturated absorption resonance occurring between the $S_{1/2}, F = 2$ to $P_{3/2}, F = 1$ transition and the $S_{1/2}, F = 2$ to $P_{3/2}, F = 3$ transition. Under this condition, the laser is 212 MHz red detuned from the $S_{1/2}, F = 2$ to $P_{3/2}, F = 3$ transition [30]. The part of the laser output to be used for producing the MOT and molasses cooling is split into six separate beams (in **Figure 8**(a), we have shown explicitly only one of these six beams). The beam for each MOT is sent through an upshifting AOM in the double-pass configuration, operating at a frequency of 100 MHz, so that each MOT beam is detuned by 12 MHz below the $S_{1/2}, F = 2$ to $P_{3/2}, F = 3$ transition. The MOT beams turn into molasses beams when each of the AOMs is tuned to 60 MHz [10]. In this way, the detuning of the molasses beams is increased to 92 MHz and the intensity is also significantly reduced because the modulation frequency is far from the central frequency of the AOMs. The double-pass configuration ensures the stability of the alignment when the modulation frequency for the AOMs varies. Note that six different AOMs are needed in order to be able to generate the moving molasses for the launching process.

The repump frequency is produced by modulating the primary MOT beam (before it is split six-ways) with an EOM driven at a modulation frequency of 6.623 GHz. For this choice of the modulation frequency, the +1 order sideband of the output of the EOM becomes resonant to the $S_{1/2}, F = 1$ to $P_{3/2}, F = 2$ transition, as needed for repumping. It should be noted that when the molasses cooling process is activated, the detuning of the repump beam from the $S_{1/2}, F = 1$ to $P_{3/2}, F = 2$ transition is ramped up to 80 MHz. Since the rate of leakage to $S_{1/2}, F = 1$ during the molasses cooling process decreases along with the detuning, the repumping rate is still adequate [3].

The Raman beams are produced in the following way. The majority of the laser output first passes through an AOM that down-shifts the beam frequency by 500 MHz, and then passes through an EOM modulated with the Raman transition frequency, which is about 6.835 GHz plus the Doppler shift resulting from the launching process. The carrier and the +1 order sideband work as the Raman frequency components. The driving power for the EOM should be adjusted to such a value that the light shifts induced by the two Raman beams are canceled. It should be noted that although this beam is intense, for example, on the order of hundreds of mW, the Raman pulses are too short to cause photorefractive damage to the EOM [3]. The AOM serves two purposes. First, it makes it possible to turn on and off the Raman beams rapidly. Second, it produces a large common detuning for them. The PBSs enable this beam to be directed to any one of the three axes depending on the states of the variable retarders VR 1 and VR 2. Here, variable retarders instead of Pockels cells are used because a fast switch process is not necessary and variable retarders do not need to be driven with high voltages. It should be noted that for each axis of operation, all output power from the AOM is directed to only one fiber port, instead of being split into three beams. The probe beam is resonant to the $S_{1/2}, F = 2$ to $P_{3/2}, F = 3$ transition, which is produced by blue shifting a piece of the laser output by 212 MHz. The imaging beam can also be directed to any of the three axes depending on the states of the variable retarders VR 3 and VR 4.

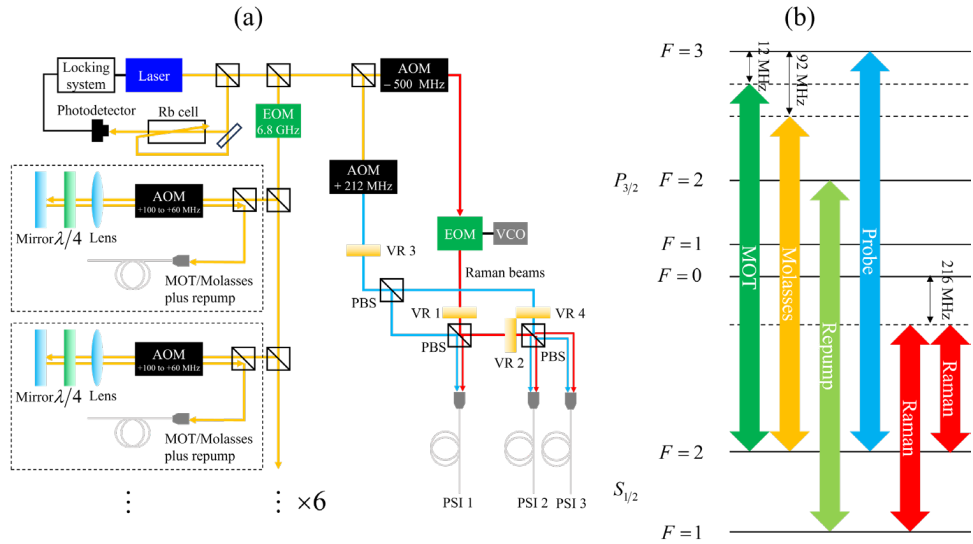

**Figure 8**: Layout of a PSI based IMU using $^{87}$Rb atom. (a) Optical system. Each of the six MOT/molasses beams is produced and controlled by an AOM optimized at a modulation frequency of 100 MHz. The modulation frequency is ramped down to 60 MHz during the process of optical molasses cooling. These beams also pass through an EOM to generate the repumping frequency component. The Raman beams (red) are produced by a 500 MHz AOM and an EOM. The imaging beam (blue) is produced by an AOM tuned to 92 MHz. The Raman beams are directed to each of the three fiber ports with PBSs and variable retarders VR1 and VR 2. The imaging beam is directed to each of the three fiber ports with variable retarders VR 3 and VR 4. (b) Frequencies of the beams used in (a).

We next describe the steps to operate each of the PSIs for the IMU. First, atoms are captured in a MOT. Then, the MOT is turned into optical molasses by turning off the MOT magnetic field and shifting the modulation frequency for the AOMs from 100 MHz to 60 MHz. Next, the frequencies of some molasses beams are reduced to produce a moving molasses field, using the protocol summarized earlier in **Table 1**. Upon the completion of the launching with the moving molasses, the molasses beams are turned off, so that the optical pumping also gets turned off. At this point, the atoms are all in the $S_{1/2}, F = 2$ hyperfine state, with populations distributed among the Zeeman sublevels defined with respect to a quantization axis that is the same as the propagation axis of the PSI being realized. This is now followed by the implementation of a state selection process. This process consists of the following steps. First, a bias magnetic field parallel to the Raman beams is turned on. Then, a Raman $\pi$ pulse is applied, with the EOM tuned to a frequency that causes a resonant, counter-propagating Raman transition (with the effective direction of $+z$) between the $F = 2, m_F = 0$ state and the $F = 1, m_F = 0$. The bias field should be strong enough to avoid excitation of Raman transitions of the other Zeeman substates. The intensity of the Raman beams for this step should be small enough so that it can select atoms in a narrow longitudinal velocity range [28, 31]. The state selection process is followed by the application of the imaging beam to blow away the remaining atoms in the $S_{1/2}, F = 2$ state.

After the state selection, the Raman pulses are applied to implement the LMT-augmented PSI protocol. For the LMT process, the propagation directions of the two Raman beams need to be swapped for some $\pi$ pulses, as discussed above. However, it should be noted that in this configuration, there already exists two pairs of counterpropagating Raman beams in opposite effective directions. Therefore, by tuning the EOM from one Doppler-broadened Raman resonant peak to the other, the propagation directions of the Raman beams can be effectively swapped. Finally, the spatial population distribution of state $S_{1/2}, F = 2$ is imaged with the imaging beam and the camera.

## 7. Discussion and Conclusion

We have presented a systematic and rigorous analysis of the sensitivity and bandwidth of accelerometry and rotation sensing with a point source atom interferometer employing large momentum transfer with molasses-launched atoms. This analysis yields a generalized expression for these sensitivities in terms of the signals measured by each pixel of a camera used for monitoring the distributed signal. We show that when the acceleration or rotation exceeds certain levels, it is possible to express the sensitivity in a closed form. Otherwise, the sensitivity can only be determined via a numerical evaluation of the general expression. We also show how the molasses-based launching process makes it possible to realize the LMT process without the need to physically change directions of the Raman pulses, thus significantly simplifying the apparatus and reducing the amount of time needed to make the measurements. These advantages become more important when this process is used for realizing an IMU. We describe an explicit scheme for a such an IMU and determine the expected sensitivity and bandwidth thereof for experimentally accessible parameters. Of course, for any specific apparatus, experimental studies need to be carried out in order to evaluate and suppress the effects of systematic sources of error.

**Funding.** National Aeronautics and Space Administration (80NSSC20C0161); U.S. Department of Defense (W911NF202076); Office of Naval Research (N00014-19-1-2181)

**Disclosures.** The authors declare no conflicts of interests.

**Data availability.** Data underlying the results presented in this paper are not publicly available at this time but may be obtained from the authors upon reasonable request.

**Appendix**

Here we derive the phase shift, due to both acceleration and rotation, for a light-pulse atom interferometer augmented by large momentum transfer (LMT). Consider first the acceleration-induced phase shift. To determine this shift in the presence of LMT, it is convenient to review briefly the derivation for the conventional $\pi/2$-$\pi$-$\pi/2$ LAI, which only includes pulses $A_0$, $B_0$, and $C_0$ in **Figure 7**. The phase shift can be expressed as

$$\phi_0 = \phi_{A_0} + \phi_{C_0} - \phi_{B_0}^{\uparrow} - \phi_{B_0}^{\downarrow} = \left( \phi_{C_0} - \frac{\phi_{B_0}^{\uparrow} + \phi_{B_0}^{\downarrow}}{2} \right) - \left( \frac{\phi_{B_0}^{\uparrow} + \phi_{B_0}^{\downarrow}}{2} - \phi_{A_0} \right) \quad \text{(A1)}$$

where $\phi_{A_0}$ denotes the relative phase between the two Raman beams at pulse $A_0$, and $\phi_{B_0}^{\uparrow(\downarrow)}$ denotes this relative phase at the upper (lower) arm at pulse $B_0$. By defining the Raman beam relative phase at the mean position at $B_0$, namely $\phi_{B_0} \equiv \left( \phi_{B_0}^{\uparrow} + \phi_{B_0}^{\downarrow} \right)/2$, this phase shift can be simplified as

$$\begin{aligned}\phi_0 &= \left( \phi_{C_0} - \phi_{B_0} \right) - \left( \phi_{B_0} - \phi_{A_0} \right) = \mathbf{k}_{\text{eff}} \cdot \left( \Delta \mathbf{r}_{C_0,B_0} - \Delta \mathbf{r}_{B_0,A_0} \right) \\ &= \mathbf{k}_{\text{eff}} \cdot \left( \overline{\mathbf{v}}_{C_0,B_0} - \overline{\mathbf{v}}_{B_0,A_0} \right) T = \mathbf{k}_{\text{eff}} \cdot \left( \mathbf{a} T \right) T = \mathbf{k}_{\text{eff}} \cdot \mathbf{a} T^2 \end{aligned} \quad \text{(A2)}$$

For simplicity, in what follows, the displacement or velocity of the mean position of the atoms are referred to as the displacement or velocity of the atoms. In Eq. (A2), $\Delta \mathbf{r}_{C_0,B_0}$, for example, denotes the displacement of the atoms between pulse $C_0$ and $B_0$, $\overline{\mathbf{v}}_{C_0,B_0}$ denotes the average velocity of the atoms in this interval, $\mathbf{k}_{\text{eff}}$ is the effective wavenumber of the counterpropagating Raman beams, whose absolute value is approximately $2k$, and $\mathbf{a}$ is the linear acceleration. It should be noted that the relation $\overline{\mathbf{v}}_{C_0,B_0} - \overline{\mathbf{v}}_{B_0,A_0} = \mathbf{a} T$ is used in the second to last step of Eq. (A2). In the case of LMT, the four additional pulses $A_1$, $B_{-1}$, $B_1$, and $C_{-1}$ yield an additional phase shift, given by the following expression:

$$\begin{aligned}\phi_1 &= 2\left[\left(\phi_{C_{-1}}-\phi_{B_{-1}}\right)-\left(\phi_{B_1}-\phi_{A_1}\right)\right] = 2\mathbf{k}_{\text{eff}} \cdot \left(\Delta\mathbf{r}_{C_{-1},B_{-1}}-\Delta\mathbf{r}_{B_1,A_1}\right) \\ &= 2\mathbf{k}_{\text{eff}} \cdot \left(\overline{\mathbf{v}}_{C_{-1},B_{-1}}-\overline{\mathbf{v}}_{B_1,A_1}\right)T_1 = 2\mathbf{k}_{\text{eff}} \cdot (\mathbf{a}T)T_1 = 2\mathbf{k}_{\text{eff}} \cdot \mathbf{a}TT_1\end{aligned} \quad (A3)$$

For the case of an LMT process with an arbitrary order $n$, the expression of the total phase shift can be easily generalized to be $\phi = \mathbf{k}_{\text{eff}} \cdot \mathbf{a}T\left(T+2\sum_{j=1}^{n}T_j\right)$.

Consider next the rotation-induced phase shift, which comes from the rotation of the vector $\mathbf{k}_{\text{eff}}$. The shift resulting from the original three pulses $A_0$, $B_0$, $C_0$ still equals the difference between $\phi_{C_0}-\phi_{B_0}$ and $\phi_{B_0}-\phi_{A_0}$. The phase $\phi_{C_0}-\phi_{B_0}$ is calculated to be

$$\begin{aligned}\phi_{C_0}-\phi_{B_0} &= \mathbf{k}_{C_0}\cdot\mathbf{r}_{C_0}-\mathbf{k}_{B_0}\cdot\mathbf{r}_{B_0} \\ &= \mathbf{k}_{C_0}\cdot\left(\frac{\mathbf{r}_{C_0}+\mathbf{r}_{B_0}}{2}+\frac{\mathbf{r}_{C_0}-\mathbf{r}_{B_0}}{2}\right)-\mathbf{k}_{B_0}\cdot\left(\frac{\mathbf{r}_{C_0}+\mathbf{r}_{B_0}}{2}-\frac{\mathbf{r}_{C_0}-\mathbf{r}_{B_0}}{2}\right) \\ &\equiv \mathbf{k}_{C_0}\cdot\left(\overline{\mathbf{r}}_{C_0,B_0}+\Delta\mathbf{r}_{C_0,B_0}\right)-\mathbf{k}_{B_0}\cdot\left(\overline{\mathbf{r}}_{C_0,B_0}-\Delta\mathbf{r}_{C_0,B_0}\right) \\ &= \left(\mathbf{k}_{C_0}-\mathbf{k}_{B_0}\right)\cdot\overline{\mathbf{r}}_{C_0,B_0}+\left(\mathbf{k}_{C_0}+\mathbf{k}_{B_0}\right)\cdot\Delta\mathbf{r}_{C_0,B_0} \\ &\equiv \Delta\mathbf{k}_{C_0,B_0}\cdot\overline{\mathbf{r}}_{C_0,B_0}+2\overline{\mathbf{k}}_{C_0,B_0}\cdot\Delta\mathbf{r}_{C_0,B_0}\end{aligned} \quad (A4)$$

Similarly, it can be shown that $\phi_{B_0}-\phi_{A_0} = \Delta\mathbf{k}_{B_0,A_0}\cdot\overline{\mathbf{r}}_{B_0,A_0}+2\overline{\mathbf{k}}_{B_0,A_0}\cdot\Delta\mathbf{r}_{B_0,A_0}$. Noting that $\Delta\mathbf{k}_{C_0,B_0}=\Delta\mathbf{k}_{B_0,A_0}\approx\mathbf{k}_{\text{eff}}\times\mathbf{\Omega}T$ in the limit $\Omega T\ll 1$, and that $\Delta\mathbf{r}_{C_{-j},B_j}=\Delta\mathbf{r}_{B_{-j},A_j}=\mathbf{r}/2$, where $\mathbf{\Omega}$ is the angular velocity and $\mathbf{r}$ is the displacement of the atoms from pulse $A_0$ to pulse $C_0$, the net phase shift resulting from the four pulses can be expressed as

$$\begin{aligned}\phi_0 &= \left(\phi_{C_0}-\phi_{B_0}\right)-\left(\phi_{B_0}-\phi_{A_0}\right) \\ &\approx \left(\mathbf{k}_{\text{eff}}\times\mathbf{\Omega}T\right)\cdot\left(\overline{\mathbf{r}}_{C_0,B_0}-\overline{\mathbf{r}}_{B_0,A_0}\right)+\left(\overline{\mathbf{k}}_{C_0,B_0}-\overline{\mathbf{k}}_{B_0,A_0}\right)\cdot\frac{\mathbf{r}}{2} \\ &\approx \left(\mathbf{k}_{\text{eff}}\times\mathbf{\Omega}T\right)\cdot\frac{\mathbf{r}}{2}+\left(\mathbf{k}_{\text{eff}}\times\mathbf{\Omega}T\right)\cdot\frac{\mathbf{r}}{2} = \left(\mathbf{k}_{\text{eff}}\times\mathbf{\Omega}T\right)\cdot\mathbf{r}\end{aligned} \quad (A5)$$

Following the same steps, it can also be shown that the rotation-induced phase shift resulting from $A_j$, $B_{-j}$, $B_j$, and $C_{-j}$ is $2\left(\mathbf{k}_{\text{eff}}\times\mathbf{\Omega}T_j\right)\cdot\mathbf{r}$. Summing up the phase shifts resulting from all pulses, the total phase shift is calculated to be

$$\phi = \left(\mathbf{k}_{\text{eff}} \times \mathbf{\Omega}\right) \cdot \mathbf{r}\left(T + 2\sum\nolimits_{j=1}^{n} T_j\right).$$